\documentclass[acmsmall,screen,nonacm]{acmart}
\usepackage{tabularx}
\usepackage{tikz}

\setcopyright{none}
\title[When Does a Cross-Device Run Justify PASS?]{When Does a Cross-Device Run Justify PASS?\texorpdfstring{\\}{ }
Obligation-Level Evidence Qualification under Partial Observability}

\author{Zimu Liu}
\email{liuzimu.lzm@antgroup.com}
\affiliation{\institution{Ant Group}\country{China}}
\author{Xiaoxi Guo}
\email{xiaoxi.gxx@antgroup.com}
\affiliation{\institution{Ant Group}\country{China}}
\author{Xiaolu Tang}
\email{lance.txl@antgroup.com}
\affiliation{\institution{Ant Group}\country{China}}
\author{Liang Yu}
\email{yuriy.yl@antgroup.com}
\affiliation{\institution{Ant Group}\country{China}}

\authorsaddresses{Authors' Contact Information:
Zimu Liu, \mbox{\nolinkurl{liuzimu.lzm@antgroup.com}};
Xiaoxi Guo, \mbox{\nolinkurl{xiaoxi.gxx@antgroup.com}};
Xiaolu Tang, \mbox{\nolinkurl{lance.txl@antgroup.com}};
Liang Yu, \mbox{\nolinkurl{yuriy.yl@antgroup.com}}.
All authors are with Ant Group, China}

\begin{document}
\raggedbottom
\begin{abstract}
A cross-device test can complete every command yet judge a business claim from
evidence belonging to another device or attempt. We introduce an executable
evidence contract that makes support qualification an explicit stage before
predicate evaluation. It binds each obligation to admissible subjects,
episodes, objects and observation scopes, and records unresolved support rather
than treating an apparently successful observation as proof.

We evaluate two implemented parts of the contract. A prospectively specified,
12-seed controlled study isolates association under reused identifiers across
2,304 claims. Given identical records and business policy,
Atomic-Facts Flat produces 160 false PASS and 76.30\% PASS recall; supplying
computed associations reduces false PASS to five and increases PASS recall
to 99.09\%.
Deterministic Closure produces zero false PASS and 100.00\% PASS recall,
improving on Atomic Flat in 11 seeds with one tie (two-sided sign
$p=0.00098$). Most of the benefit comes from explicit association rather than
graph notation.

A separate prospective field characterization spans four workflows,
40 physical episodes and 80 claims. Against frozen reference labels, the
physical-evidence verifier attains 95.00\% accuracy and 100.00\% PASS recall (27/27),
with three false PASS judgments.
The field results are descriptive: a policy-input mismatch prevents the planned
matched-policy comparison, and post-hoc sensitivity does not establish superiority
over a policy-informed textual baseline. Together, the studies characterize
qualification on physical evidence and isolate the value of explicit
association under a controlled matching contract.
\end{abstract}

\maketitle

\section{Introduction}\label{introduction}

Consider a test in which a terminal prepares an NFC interaction, a mechanical
arm brings a phone near it, and the phone should open an application page.
Which observation establishes success? An actuator acknowledgement establishes
neither application launch nor payment. A terminal screenshot is not a phone
screenshot, and a successful command cannot substitute for either. In the
recorded diagnostic in Figure~\ref{fig:running-evidence}, the phone obligation
was answered using the terminal image. All retained actions completed, yet
the evidence used for the local verdict concerned the wrong participant.
Another assertion made the aggregate test fail, hiding this local error in
the case-level outcome.

Testing environments with centralized state checks, including
AndroidWorld~\cite{rawles2024androidworld}, offer useful reference oracles within
their supported tasks. Cross-device physical workflows can instead expose only
partial interfaces: command receipts, per-device images, resource state, and
selected service responses. Our problem is not that state-based verification
is universally unavailable, but that these interfaces do not automatically
establish the associations required by a particular business claim.

We study an \emph{executable evidence contract}: a specification connecting
actions to required observations and versioned checks. It distinguishes the
action actor from the observation subject, binds evidence to an episode and
declared object, and limits each conclusion to its observation scope.
We call the physical-evidence implementation \emph{Graph} and the controlled
association solver \emph{Closure}; they evaluate different contract subsets.
Both implement support qualification before predicate evaluation.

We call evidence selection \emph{support qualification}: deciding which
observations may support the current claim. Even an authentic success record
can belong to another device or attempt. Distributed oracles and partial
monitoring provide relevant foundations; our focus is making this selection
explicit for heterogeneous test artifacts.

This distinction yields two oracle stages: qualify which evidence may support
an obligation, then evaluate the predicate on that support. A true predicate
on the wrong observation is not a successful test of the obligation.
The method selects support, checks its relation and predicate, and records
sources or unresolved gaps. With a unique observation, actor and step binding
can repair the wrong-image defect. With reused identifiers, a same-key match
may be inadmissible because other events consume that candidate. Our controlled
study isolates this second problem. The contribution is the executable contract
and its enforcement; graphs are one representation of support paths.

We first test the association mechanism under controlled identifier reuse.
Across 12 independent seeds and 2,304 claims, Atomic-Facts Flat receives all
records and the full policy but makes 160 false PASS among 1,532 non-PASS
claims. Closure makes none, improving in 11 seeds with one tie (two-sided sign
$p=0.00098$). Providing the computed paths to Serialized-Relations Flat reduces
its false PASS to five. This control locates the dominant benefit in making
associations explicit before judgment.

We then characterize qualification on physical workflows. Across 40
prospectively frozen episodes and 80 claims, Graph attains 95.00\% accuracy
and 100.00\% PASS recall, with three false PASS. The field study identifies
concrete binding and enforcement failures. Its comparator-input deviation
and policy-transfer sensitivity are reported in
Section~\ref{sec:natural-study}.

This paper makes three contributions:
\begin{itemize}
\item An executable evidence contract separating support qualification from
predicate evaluation, with explicit subject, episode and claim scope.
\item Two implementations: Graph binds physical observations to claims;
Closure builds contract-constrained associations across records before
evaluating their outcomes.
\item A pre-specified 12-seed mechanism study identifying the role of explicit
association, and a prospective 40-episode field study characterizing
qualification on physical evidence.
\end{itemize}

\section{Motivating Failure}\label{motivating-cases}

\subsection{A Completed Execution with an Unsupported Verdict}
\label{a-phone-assertion-answered-with-terminal-evidence}

\begin{figure}[tb]
\centering
\includegraphics[width=\linewidth]{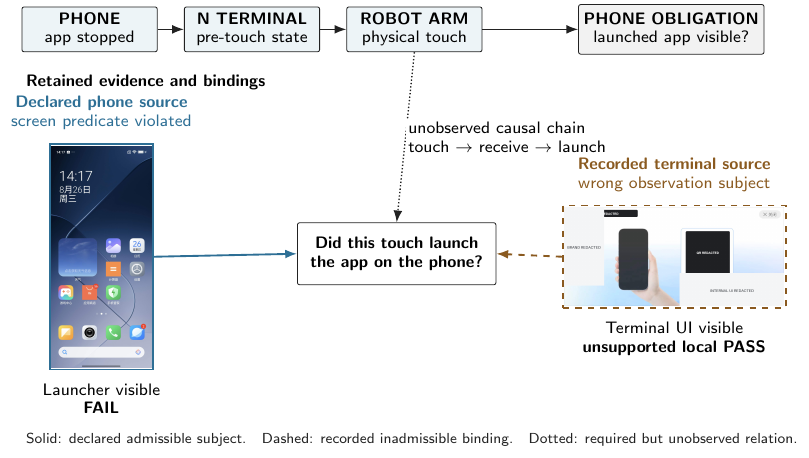}
\caption{A real cross-device episode exposes the execution--verification gap.
All retained actions completed, but a phone-side obligation consumed a later
terminal image and returned a local PASS. Binding the obligation to the phone
observation instead exposes the launcher and rejects the claim. Dashed denotes
the wrong source; dotted denotes the unverified handoff. The aggregate case
eventually failed for another reason; the figure isolates an obligation-level
unsupported PASS rather than estimating its prevalence.}
\Description{A workflow leads from a stopped phone app through an NFC terminal
and robot-arm touch to a phone-launch obligation. Below it, the phone screenshot
shows the Android launcher while the N-terminal screenshot shows its own
interface. A solid arrow binds the obligation to the declared phone source and
yields failure. A dashed arrow depicts the recorded terminal source and local
pass. A dotted line marks the missing cross-device causal link.}
\label{fig:running-evidence}
\end{figure}

The recorded episode contains a phone reset, a terminal
screenshot before interaction, a physical touch, and screenshots of the
phone and terminal afterward. The enabled plan retains source step
numbers 1, 2, 3, 6, and 7. Step 6 requires the phone to show the
application page opened by the NFC event. In the saved judgment,
however, its image URL matches the terminal screenshot from step 7. This
association appears in all four inspected device groups.

The groups illustrate different consequences of the same defect. In
group 116, the phone image shows the Android launcher, but the phone
assertion receives PASS using the terminal image. In group 119, a
relevant application page is present in the phone image, but the judge
does not use that image and returns FAIL for the phone assertion. That
image alone does not establish complete workflow success or absence of
payment effects. In group 120, acquisition of the phone screenshot
fails, yet the phone assertion is still evaluated against the terminal
image. The missing phone observation and the business state of the phone
are different questions.

A second multi-actor diagnostic reproduces the association
problem in a chain that also includes POS initialization and terminal
configuration. Across three inspected groups, all four visual assertions
use the final terminal screenshot, including the assertion explicitly
assigned to the phone. These observations extend the reproduction across
two task chains, but both use the same testing infrastructure and some
device groups overlap. They do not establish prevalence across
independent systems. A keyed selector is an immediate and necessary
comparator for any proposed solution.

\subsection{Two Further Forms of the Same Gap}
\label{a-requirement-present-in-text-but-absent-from-checking}

A display test requests a QR code and then a screenshot. Its second
expectation states, ``The final screenshot contains a visible QR code.''
The saved validation records a step-execution check with the actual
result ``screenshot succeeded'' rather than an image-content check. The
final image shows the terminal application's menu without a visible QR
code.

The judgment must be read in full. It contains original\_status=BLOCKED,
status=PASS, and review\_required=true, and explains that some
expectations lack passing assertions. This is not an unqualified
automated acceptance. It is evidence that a PASS status can coexist with
an unevaluated business requirement. An oracle should preserve the
distinction between successful acquisition, actual predicate evaluation,
and unresolved review. This case involves a single terminal and serves
as a mechanism control, not as another cross-device workflow.

\paragraph{Evidence with limited scope.}
\label{an-observation-with-a-limited-business-meaning}

In the audited amount-touch round, four payment calls have
successful outer responses and successful inner queries. Two queried
orders report TRADE\_SUCCESS with a paid amount of 0.01, while two
report WAIT\_BUYER\_PAY with a paid amount of 0.00. The complete
responses already distinguish these states; only the interface-success
projection is shared. The cases do not demonstrate identical complete
observations with different business outcomes.

The archive also limits what can be linked. The payment response
contains order-query parameters, but the independent outbound payment
request is not preserved in the saved envelope. Query-to-response and
payment-to-refund order equality can be checked; the link from the
received dynamic code to the dispatched payment request cannot be
independently certified from these artifacts. Likewise, the pending
orders cannot be called eventual failures without an appropriate
deadline and sufficient subsequent observation. This case motivates
explicit scope for both identity and time, rather than a stronger
inference from an existing status field.

\subsection{Generalized Requirements}\label{requirements-derived-from-the-cases}

\begin{table}[tb]
\caption{Evidence breaks derived from the motivating cases. Categories may
co-occur and do not estimate prevalence.}
\label{tab:evidence-breaks}
\centering
\footnotesize
\renewcommand{\arraystretch}{1.12}
\begin{tabularx}{\linewidth}{@{}>{\raggedright\arraybackslash}p{0.13\linewidth}
>{\raggedright\arraybackslash}X >{\raggedright\arraybackslash}X
>{\raggedright\arraybackslash}X@{}}
\toprule
Break & Diagnostic question & Observed defect & Required gate \\
\midrule
Coverage & Was the required predicate checked? & Requirement exists in text,
but no executable check evaluates it. & Bind every obligation to a versioned
evaluator. \\
Association & Does the evidence belong to the required actor and episode? & A
terminal image is used for a phone assertion. & Match actor, step, episode, and
object keys. \\
Scope & What exact claim can the observation establish? & A valid state is used
to infer a broader, unobserved handoff. & Check authority, identity, time,
completeness, and causal reach. \\
\bottomrule
\end{tabularx}
\end{table}

Table~\ref{tab:evidence-breaks} derives requirements, not an exhaustive taxonomy.
Correcting an evidence path need not improve aggregate accuracy; evaluation
must retain both assertion-level behavior and workflow outcomes.

\section{Cross-Device Testing Model}\label{problem-boundary-and-candidate-design}

We model a cross-device test as a versioned episode $e$ containing actors,
actions, obligations, and observations. An \emph{actor} is a logical role such
as a phone, terminal, service, or physical actuator. An \emph{action} is a
capability-bound operation performed by one actor. An \emph{obligation} is a
property that must be established for the test to pass. An \emph{observation}
is a retained fact or artifact about a subject. The actor performing an action
and the subject exhibiting its effect may differ.

We use \emph{claim} for a concrete assertion evaluated in an episode and
\emph{obligation} for the requirement it is meant to establish.
\emph{Support qualification} asks whether selected evidence can support
that claim; \emph{association} establishes which records belong together.
\emph{Predicate evaluation} asks whether the evidence shows the required
condition. \emph{Scope} limits the conclusion, for example to a visible
success page rather than confirmed settlement.

An executable evidence contract $C_e$ contains the actions required to exercise
episode $e$ and a nonempty set $O_e$ of obligations. In the field study, contracts,
checks, and analysis are frozen before execution; earlier exposed cases inform
design but are excluded from its score. Each obligation specifies a trigger,
predicate, admissible subjects
and channels, association keys, temporal conditions, and observation scope.
The execution runtime resolves actor capabilities and records action receipts
and observations. The verifier determines whether those observations are
eligible to support each obligation.

The central safety objective is \emph{no unsupported PASS}: a conforming
verifier should return PASS only when every applicable necessary obligation has
qualified positive support. Qualification requires (1) an executable check for the predicate,
(2) binding to the observation subject, episode and claim-required object/action,
(3) sufficient
observation scope, and (4) satisfaction of required temporal conditions. A
phone image can establish a visible page without proving which physical action
caused it; a service query can establish an object's current state without
linking it to an unrecorded initiating request. The field study measures rather than
assumes conformance. Unsupported PASS violates the contract; a false PASS
against reference labels is a disagreement requiring contract analysis.

Never returning PASS is trivially safe. We therefore pair false PASS with
supported-PASS utility, coverage and attribution. The field study characterizes
frozen workflows; the construction isolates association. Six-class verdicts
retain support or unresolved evidence. Model transport failures are unavailable
predictions, not business labels.

Our scope is testing with heterogeneous device and service interfaces when a
trusted global-state oracle is unavailable. We do not claim to authenticate a
malicious evidence source, reconstruct an unobserved causal link, or diagnose
the root cause inside the system under test. Source binding also does not
certify the meaning of a visual predicate: a correctly acquired image can
still be misinterpreted. Required business meanings must remain common to
comparators; withholding them tests task knowledge, not representation.
The method makes these evidence boundaries explicit and auditable.

\section{Executable Evidence Contracts}\label{sec:method}

The evaluated core has two stages. First, bind observations to the claim's
subject and episode and construct the associations required by its contract.
Second, evaluate the predicate on that selected support and retain the
support references or unresolved gaps. Table~\ref{tab:design-rationale} maps
this core to Graph's physical-evidence checks and Closure's controlled joins.

\paragraph{Beyond independent field selection.}
A unique phone image can be selected by subject, episode and step. Reused keys
create a different problem: several records may match those local fields,
while the contract allows a predecessor to support only one successor.
Selecting each claim's most plausible success path independently can then
reuse evidence already assigned elsewhere. Closure builds the stage-wide
assignments before inspecting outcome predicates. In
Figure~\ref{fig:association-conflict}, this assigns the success tail to a
competing touch and the declined path to the queried touch. The required
operation is constrained association; it can be implemented with keyed maps
or relational queries as well as graph paths.

The sections below specify the evaluated bindings, associations and checks.
Broader applicability, deadline and cleanup semantics are summarized at the
end; they are not jointly validated by these implementations.

\begin{table}[H]
\caption{Contract elements and evaluated instantiations.}
\label{tab:design-rationale}
\centering\footnotesize
\begin{tabularx}{\linewidth}{@{}>{\raggedright\arraybackslash}p{0.20\linewidth}>{\raggedright\arraybackslash}p{0.27\linewidth}>{\raggedright\arraybackslash}X@{}}
\toprule
Element & Natural implementation & Controlled implementation \\
\midrule
Observation & Bound image/receipt and extracted visual fact & Typed input
records; perception and physical-source binding not tested \\
Association & Episode, step and observation subject & Four global keyed maps
with latest-unused predecessor consumption \\
Predicate & Workflow-specific action and visible-state checks & Amount,
backend approval/decline and phone success/home along the assigned path \\
Evidence/status gaps & Missing/ambiguous observations and classified execution
blocks & Missing hops, amount mismatch or indeterminate path state \\
\bottomrule
\end{tabularx}
\end{table}

\subsection{Actions, Obligations, and Observation Subjects}
Let $A$ and $O$ denote actions and primary-business obligations; cleanup is
separate. Each action declares an actor role, a required capability,
parameters, ordering dependencies, and a failure policy. The runtime binds
roles to available implementations but may not silently replace a missing
capability with a semantically different action.

For each obligation, the specification records a trigger, a required predicate,
the permitted observation subjects and channels, an association rule, and any
deadline. For a prospective evaluation, requirements must be established before
inspecting its outcomes.
A missing mandatory field is a specification error; an empty obligation set
must not yield a vacuous business PASS.

An action actor and an observation subject are separate fields. For example,
a terminal configuration command may precede a physical interaction whose
required effect is an application page on the phone. The command receipt,
actuator receipt, and phone image support different statements. A cross-device
obligation must declare how the receiving observation belongs to the triggering
episode. It cannot inherit the action actor as its observation subject merely
because both occur in the same authored step.

\subsection{Evidence Obligations and Decision Records}
A check contract connects an obligation to a versioned evaluator and its
admissible inputs. It specifies what constitutes positive evidence, what
constitutes a counterexample, and which inputs require abstention. The
requirement identifier is a reference to this contract, not a proof that the
evaluator checks the requirement. For example, an evaluator that only tests
whether an image file was acquired cannot cover a visible-page predicate,
even if its output names the correct requirement.

\paragraph{Implemented example (E031).}
The frozen \texttt{phone\_target\_state\_visible} claim consumes steps
1, 2, 4 and 5: home, phone capture, application start and phone capture.
The claim binding fixes its episode, workflow, claim set and common-input
hash before prediction. A unique successful start receipt at step 4 anchors
pre/post phone observations; terminal captures are excluded by subject.
The predicate requires a readable current target-app surface after start,
with pre/post observations present. Here \texttt{visual:1} is the step-2
launcher and \texttt{visual:3} the step-5 payment-result surface. The frozen
output is PASS with these two \texttt{evidence\_refs}, the four
\texttt{dependency\_steps}, and
\texttt{capture\_binding\_refs\_verified=true}.
The wrapper validates episode/step/image binding; it does not prove capture
freshness. Missing support yields uncertainty; readable non-target evidence
yields FAIL. This post-hoc illustration establishes neither payment success
nor the visual extractor's correctness.

The implemented decision record retains the claim, verdict, source references
and dependency steps. These fields separate selection errors from predicate
errors. Contract authoring and visual interpretation are separate sources of
error; source hashes establish neither business intent nor perceptual accuracy.

\subsection{Evidence Qualification and Association}
Each observation retains a source reference, collection interval, subject,
channel, available identifiers for steps, requests, and business objects, and the scope of the
extracted fact. Derived facts retain their parser version and source locations.
Collection success and interpretation are separate records. A screenshot operation
can succeed while the requested screen predicate remains unchecked.

Association uses the declared keys and scope. A unique step/actor image is
sufficient only where the obligation actually refers to that step and actor.
An end-to-end obligation may require a request identifier, transaction key,
or a separately justified isolation condition. Timestamp proximity alone is
not an association certificate. Missing keys, multiple admissible candidates,
and disagreement about the observation subject remain explicit gaps.
In Figure~\ref{fig:running-evidence}, a keyed phone image can support the
screen-state predicate while the touch-to-launch obligation remains unresolved.
The two obligations require different association evidence.

The association requirement must not exceed the declared claim. A phone-page
predicate at a named checkpoint does not require a backend transaction key.
An ordered, single-attempt UI-consistency claim can compare the terminal amount,
phone outcome and terminal outcome within its stated isolation assumptions,
without claiming settlement. In contrast, assigning those observations to one
of several competing transactions requires an independently justified
discriminator. A host query's matching request and response identifiers bind
that query, not an unkeyed phone or terminal image. Missing transaction identity
therefore leaves the stronger claim unresolved; it does not automatically
invalidate a correctly scoped page observation.

Hash checks detect changes relative to a stored manifest; they do not authenticate
the device or validate an image interpretation. Likewise, a query receipt can
establish the state of its queried object at that time without establishing
that the object belongs to an earlier payment request. Evidence rejected for
one obligation can remain relevant to another, but must not be substituted
without checking the latter's contract.

Association validity and predicate truth are separate decisions. A bound
launcher image contradicts application visibility; an unrelated success image
is inadmissible regardless of its content. Shared consumption constraints can
make a locally compatible path inadmissible
(Figure~\ref{fig:association-conflict}). If multiple admissible
assignments imply different answers, the contract requires abstention. The
latest-unused rule is a declared construction policy, not a rule recovered
from physical execution or justified by receipt arrival order.

\paragraph{Implemented controlled association.}
For each adjacent record stage, let $P$ be its predecessor records and $Q$
its target records. Initialize $U=\emptyset$, then process $q\in Q$ in
increasing event time. With the join's declared key $k$ and timestamp $t$,
form $E_q$ and, when nonempty, select $p_q$:
\[
 E_q=\{p\in P\setminus U:k(p)=k(q),\ t(p)<t(q)\},\qquad
 p_q=\mathop{\arg\max}_{p\in E_q}(t(p),\mathrm{id}(p)).
\]
If $E_q$ is empty, leave the target unmatched; otherwise record
$M(p_q)=q$ and add $p_q$ to $U$. Reset $U$ for each of the four joins.
All maps are built before tracing any claim from its touch anchor.
Only then are amount and outcome predicates evaluated: association never
selects a record because its state looks successful. Missing hops or amount
mismatch yield uncertainty; a complete amount-matching path yields PASS for
backend approval and phone success, FAIL for decline or phone home, and
uncertainty otherwise. The evaluated cohort has distinct within-stage times;
the implementation's ID tie-break is not tested.

\begin{figure}[tb]
\centering
\includegraphics[width=\linewidth]{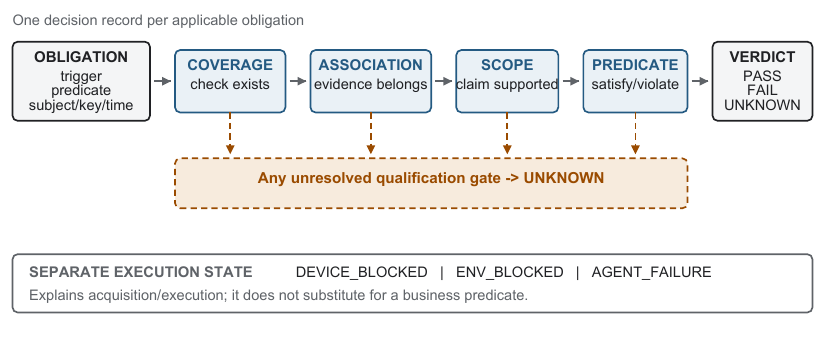}
\caption{Conceptual evidence contract: coverage, association, scope and
predicate checks qualify support for each obligation. The evaluated subsets
are listed in Table~\ref{tab:design-rationale}; applicability and deadline
semantics are extensions. The separate lane distinguishes execution state
from business evidence.}
\Description{An obligation specification containing trigger, predicate,
permitted subjects, association keys, scope, and time conditions enters four
gates: coverage, association, scope, and predicate evaluation. The gates feed
PASS, FAIL, or UNKNOWN business verdicts. A separate lane records device,
environment, and agent execution states.}
\label{fig:evidence-contract}
\end{figure}

Figure~\ref{fig:evidence-contract} operationalizes this distinction as four
gates applied to each declared obligation. A screen-state check cannot
discharge an additional obligation about which interaction triggered that
state, and execution failures remain separate from the business predicate.

For the running episode, the contract separates screen-state and
touch-to-launch obligations. The Graph admits the step-6 phone image through
its subject--step--episode path; the launcher is counterevidence, so the local
verdict is FAIL. It excludes the later terminal image on subject mismatch and
records that failed qualification. With no receive-to-launch link, the causal
obligation remains unresolved rather than being inferred from screen state.

A visible postcondition and a causal attribution are different claims.
An isolated trace can support a functional postcondition under its declared
assumptions; attributing it to a physical interaction additionally requires
evidence of the receive-to-launch link. The verifier must evaluate the declared
scope rather than infer that link from a success-looking screen.

\subsection{Verdicts and Unresolved Support}
For each claim, the verifier evaluates the predicate only on qualified
support. Positive evidence supports PASS within the declared scope;
qualified counterevidence supports FAIL. Missing or ambiguous support remains
unresolved. For example, a pending payment state is neither a success nor
necessarily a terminal failure. The implemented matching and verdict rules
are frozen before evaluation. This does not imply that the reference-label
instructions resolve every conflict between a workflow prerequisite and a
local visible-state claim; Section~\ref{sec:discussion} discusses that gap.

The natural implementation emits one of six labels: \texttt{PASS},
\texttt{FAIL}, \texttt{DEVICE\_BLOCKED}, \texttt{ENV\_BLOCKED},
\texttt{AGENT\_FAILURE}, or \texttt{ORACLE\_UNCERTAIN}. A verified execution
block may determine the label before visual evaluation. This single label
summarizes execution and evidence status; it does not encode all combinations
of those states. Invalid inputs are rejected, and a model response that is
unavailable is not classified as a business failure.

Closure uses PASS, FAIL and ORACLE\_UNCERTAIN for its assigned paths.
Both evaluations score individual claims. This preserves errors that an
aggregate case-level failure could conceal, as in the motivating episode.

\paragraph{Contract extensions outside the evaluated core.}
A full case-level contract would also need applicability, completeness and
deadline rules before aggregating its necessary obligations. An empty set or
unknown applicability cannot justify PASS, and absence requires a complete
observation interval before it establishes failure. These features are not
jointly evaluated here. Resource cleanup is separate from business judgment;
the field cohort does not evaluate refunds or ledger settlement. Additional
queries and physical retries require separate evaluation, since a new action
can change the episode rather than merely reveal its prior state.

\section{Implementation and Study Boundaries}\label{sec:implementation}

An internal platform supplies capability-bound execution and versioned bundles
of images, receipts and bindings. A visual extractor emits typed observations
for a deterministic verifier. Hashes establish stored-input identity, not
authenticity or business truth.

The field wrapper uses the six-label interface defined in
Section~\ref{sec:method}. Invalid evidence bindings produce no business
verdict and are not treated as correct non-PASS predictions. The frozen scorer
rejects invalid labels, duplicate claims, or an incomplete prediction population
instead of silently scoring a smaller subset. In the retained cohort, all
40 Graph evaluations succeeded, with zero blocked evaluations; both arms
provide valid labels for all 80 claims. Accuracy therefore uses all 80 claims,
PASS recall uses the 27 reference PASS claims, and false-PASS rate uses the
53 reference non-PASS claims.

The prospective study freezes both arms before reference-label access. Flat interprets
the same source evidence independently of the Graph extractor. The arms differ
in decision-policy input; Section~\ref{sec:evaluation} reports that deviation
and a separate post-Gold policy-transfer sensitivity. Shared provider, prompt,
extraction, and source-definition biases remain possible.

The evaluated scope follows Table~\ref{tab:design-rationale}: Graph checks
image-to-receipt and episode/step bindings with selected business predicates;
Closure associates typed abstract records under a specified matching rule.
Object identity and ordering are partial or controlled-only. Contract authoring,
general applicability, deadlines, source authentication and cleanup are outside
the evaluated core.

The execution-recovery prototype is an opt-in local timeout quarantine. When main-step timeout
is observed, it closes progress persistence, keeps the timeout state, requests
worker interruption, and records \texttt{execution\_unresolved}. It skips
later steps, cleanup and automatic release. The evaluated path uses actual
orchestrator, step-runner, and capability-dispatcher sources; action handlers,
mapper and release-controller boundaries record events without external I/O.
This validates a local mechanism, not remote cancellation, actual database
persistence, provider fencing or a full resource-recovery protocol.

\subsection{AI Assistance and Reference Separation}
An AI coding assistant supported protocol drafting, implementation, audit,
diagnostic analysis, and manuscript editing under author review. In the natural
study, the frozen Flat and visual-extraction calls use the requested model identifier
\texttt{gpt-5.5} with low reasoning effort; all 80 calls complete with no retry,
replacement, tool event, or Gold access. The controlled relation study uses
\texttt{gpt-5.6-sol} with high reasoning effort, frozen complete-evidence
prompts, fixed 64-claim shards, schema validation, and a pre-specified
transport-only retry rule. Every attempt is retained and the first valid
response is selected; no output is replaced based on its verdict. Provider
records do not independently attest the backend weights. Natural reference
labels come from the reviewer procedure in Section~\ref{sec:natural-study};
model-assisted pre-annotation is excluded from Gold. The synthetic reference
is produced by the frozen generator and opened after held-out predictions
are fixed. The 40 Policy-Transferred Flat sensitivity calls were made after
Gold existed. Their original one-shot outputs were retrieved from persisted
completed responses and are analyzed as post-hoc sensitivity.

\begin{table}[tb]
\caption{Recorded configuration of the two primary model studies. Dates identify
the archived protocol snapshots, not an independently verified execution clock.
Both used Codex CLI 0.144.6. Model names are requested identifiers.}
\label{tab:execution-config}
\centering\small
\begin{tabularx}{\linewidth}{@{}lXX@{}}
\toprule
Setting & Controlled association & Natural cohort \\
\midrule
Protocol snapshot & 2026-09-24, v3 & 2026-09-25, v1 \\
Requested model & \texttt{gpt-5.6-sol} & \texttt{gpt-5.5} \\
Reasoning effort & High & Low \\
Model calls & 72 valid; 75 attempts & 80 valid; 80 attempts \\
Concurrency limit & 3 workers & 5 calls \\
Retry policy & At most two transport-only retries & No retry or replacement \\
Output unit & 64-claim shard & One episode per call \\
Prompt scope & Full records and policy; computed paths in Serialized Flat & Shared source evidence; arm-specific instructions \\
\bottomrule
\end{tabularx}
\end{table}

The natural prompt projections are retained as
\path{REGISTERED_FLAT_HEADER.txt} and
\path{GRAPH_VISUAL_HEADER.txt}; the sensitivity uses
\path{POLICY_TRANSFERRED_FLAT_HEADER.txt}. Controlled prompts and their
input hashes are retained with the per-call records. These are local provenance
locations, not public download links (see Data Availability).

\section{Evaluation}\label{sec:evaluation}

\textbf{R1:} Under the declared matching contract, how does materializing
associations affect false PASS and PASS recall?
\textbf{R2:} What qualification failures and verdict errors occur in the frozen
natural workflows? R1 isolates the association mechanism with seed-level
inference; R2 characterizes qualification on physical evidence descriptively.

\paragraph{Study and method map.}
R1 compares Closure with two model judges: Atomic-Facts Flat reconstructs
associations from records, and Serialized-Relations Flat receives computed
paths. R2 evaluates Graph on physical evidence, alongside the frozen Min-Flat
comparator and a subsequent Policy-Transferred Flat sensitivity.
Both studies examine qualification before predicate evaluation, but use
different implementations and inputs; comparisons are within each study.
The supporting timeout study checks execution boundaries.

\begin{table}[htb]
\caption{Inputs and association responsibility. Comparisons use the same
source evidence within each study; policy availability differs in the
natural study. The final row is a post-Gold sensitivity.}
\label{tab:comparator-inputs}
\centering\footnotesize
\begin{tabularx}{\linewidth}{@{}>{\raggedright\arraybackslash}p{0.24\linewidth}>{\raggedright\arraybackslash}p{0.32\linewidth}>{\raggedright\arraybackslash}X@{}}
\toprule
Method & Decision knowledge & Association and judgment \\
\midrule
\multicolumn{3}{@{}l}{Controlled records (R1)} \\
Atomic-Facts Flat & Full written matching and verdict policy &
Model reconstructs paths and judges them \\
Serialized-Relations Flat & Same full policy &
Computed Closure paths supplied; model judges \\
Closure & Same policy implemented in code &
Program constructs maps and evaluates paths \\
\midrule
\multicolumn{3}{@{}l}{Physical images and receipts (R2)} \\
Graph & Implemented verifier policy &
Model extracts facts; program binds and judges \\
Min-Flat & Claim criteria and policy hash; full policy absent &
Model interprets serialized evidence and judges \\
Policy-Transferred Flat & Full written policy and four verifier sources &
Model interprets evidence and judges; no graph edges supplied \\
\bottomrule
\end{tabularx}
\end{table}

\subsection{Controlled Relation-Reconstruction Study}\label{sec:controlled-study}
We stress association under controlled identifier reuse, specified independently
of the unopened field-study outcome. Each claim anchors a touch record and an
expected amount. Four joins connect touch, terminal intent, backend outcome,
order--session bridge, and phone UI through touch, payment, order, and session
keys, respectively. Each key type draws from eight reused opaque tokens.

Closure applies the four global latest-unused maps defined in
Section~\ref{sec:method}, then evaluates each claim's assigned path. This
matching contract is a controlled construction, not a validated production
rule; physical perception and source binding are outside this study.
Its verdict space is PASS, FAIL and ORACLE\_UNCERTAIN; macro-F1 averages
these three classes, unlike the field study's fixed six-class metric.
The two macro-F1 scales are not directly comparable.

\begin{figure}[htb]
\centering
\includegraphics[width=\linewidth]{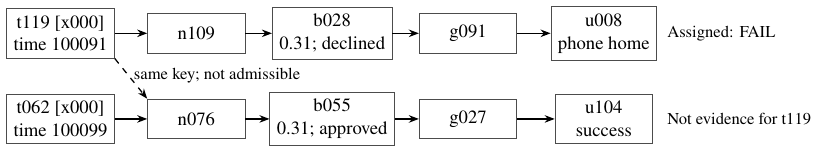}
\caption{Controlled claim c016 expects 0.31 for t119. Both touches share key
x000. Solid paths respect latest-unused assignment; the dashed candidate does
not. The success tail belongs to t062; t119's assigned path is declined.}
\Description{Two touches share a key and amount. The success path belongs to
the later touch; the claim's own path contains declined and phone-home states.}
\label{fig:association-conflict}
\end{figure}

\paragraph{Worked association conflict.}
Figure~\ref{fig:association-conflict} shows a post-hoc error example,
claim \texttt{c016} in the first retained scenario. Atomic Flat returns PASS;
Serialized Flat and Closure return FAIL. The figure contrasts the admissible
path with a competing success path.

\paragraph{Design and comparators.}
We pre-specified 12 fresh seeds, four each at densities 128, 192 and 256:
2,304 claims. Atomic-Facts Flat receives all records, policy, claims and
identities; Serialized-Relations Flat also receives Closure's computed paths.
Closure reconstructs associations, rejects globally inconsistent paths and
does not copy reference labels. Both tool-free model arms receive the full
scenario and return 64-claim shards, separating response length from association
workload. The comparison tests direct reconstruction, not code-assisted agents.

Both arms request \texttt{gpt-5.6-sol} with high reasoning via the same official
client. Three workers execute 72 calls with 1,200-second limits and up to two
transport-only retries; only the first valid tool-free completion is admitted.
Protocol, dependencies, inputs and analysis are frozen before dispatch;
outputs are hash-verified and frozen before reference access. All 12 scenarios
complete in 75 attempts: 72 valid calls, three retries and no replacements.

\begin{table}[htb]
\caption{Independent-seed synthetic study. All methods cover 2,304 claims,
including 1,532 non-PASS and 772 PASS. Rates pool claims; inference uses seeds.}
\label{tab:relation-replication}
\centering\small
\begin{tabular}{@{}lrrr@{}}
\toprule
Method & False PASS & PASS recall & Macro-F1 \\
\midrule
Atomic-Facts Flat & 160/1,532 & 76.30\% & 78.08\% \\
Serialized-Relations Flat & 5/1,532 & 99.09\% & 99.18\% \\
Contract Closure & 0/1,532 & 100.00\% & 100.00\% \\
\bottomrule
\end{tabular}
\parbox{\linewidth}{\footnotesize A post-hoc SQLite control matches Closure
on all 2,304 verdicts. This is implementation agreement under the shared
contract, not independent oracle validation or a pre-specified comparison.}
\end{table}

\paragraph{Paired results.}
For the pre-specified primary comparison, Closure reduces false-PASS rate relative
to Atomic Flat in 11 seeds, with one tie and no reversals (two-sided exact sign
$p=0.0009766$).
The equal-seed mean reduction is 9.73 percentage points with a seed-bootstrap
95\% interval of [5.24, 15.34]; corresponding PASS-recall and macro-F1 gains
are 20.61 [10.64, 31.35] and 19.66 [11.81, 27.58] points.
These equal-seed estimands differ from the pooled rates in
Table~\ref{tab:relation-replication} because scenario sizes differ.

\begin{figure}[htb]
\centering
\includegraphics[width=\linewidth]{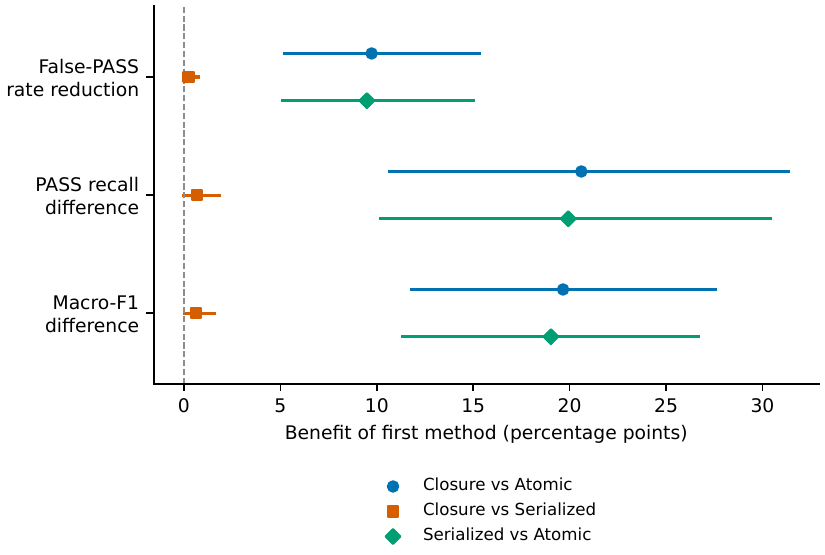}
\caption{Equal-seed paired mean effects and 95\% seed-bootstrap intervals
(10,000 draws). Only Closure versus Atomic is the pre-specified primary test;
the two other contrasts are descriptive secondary comparisons.}
\Description{Paired mean effects across twelve seeds. Closure and Serialized
Flat both improve substantially over Atomic Flat. The Closure versus Serialized
false-PASS reduction is small and its interval includes zero.}
\label{fig:relation-replication}
\end{figure}

The comparison separates association reconstruction from predicate judgment
with supplied associations. Serialized Flat removes most of Atomic Flat's
errors. Closure's five remaining
false-PASS improvements over Serialized Flat occur in one seed, with 11 ties
(descriptive sign $p=1$); its equal-seed rate reduction is 0.25 points
[0.00, 0.74]. The result favors an explicit association stage, not a graph-only
verdict engine. Closure superiority over equally informed text is not
established. Perfect Closure is contract conformance,
not evidence of perfect natural perception or association. All reference
non-PASS claims have competing success tails, so candidate presence does not
explain model errors. Clean-room replay reproduces the statistics.

\paragraph{Post-hoc relational control.}
A separately implemented SQLite control selects latest-unused predecessors
with anti-joins, then applies joins and a verdict predicate. It imports neither
Closure nor the generator and reads no reference labels during prediction.
All 2,304 verdicts across the 12 scenarios match frozen Closure. This
post-hoc implementation ablation is outside the pre-specified comparison:
deterministic association suffices here without a graph representation.
Shared contract assumptions mean agreement is not independent oracle validation.

\subsection{Natural-Workflow Characterization}\label{sec:natural-study}
We use a fresh cohort (artifact identifier: \texttt{natural-v8}) collected only
after the test specification, evidence schema, Graph
verifier, Flat prompt, six labels, and analysis plan were frozen. Earlier
field observations informed development and are excluded from this score.
The cohort contains 40 episodes and 80 judgments: ten episodes and two
claim types per workflow, with ten instances of each type
(Table~\ref{tab:workflow-claims}). Touch execution is action-only; payment
claims concern visible outcomes rather than ledger settlement. A
blocked schedule assigns episodes across four device groups and five time
blocks. Every scheduled episode remains in the denominator; none is retried,
replaced, or selected using its outcome. These heterogeneous claims include
action completion and observed business conditions; their accuracy is not
an end-to-end workflow completion rate.

\begin{table}[htb]
\caption{Natural workflow claims and frozen reference distribution.
Each row contains ten episodes and 20 claims. P/B/F/U denote PASS,
DEVICE\_BLOCKED, FAIL and ORACLE\_UNCERTAIN. The other two labels have no
reference instances.}
\label{tab:workflow-claims}
\centering\small
\begin{tabularx}{\linewidth}{@{}>{\raggedright\arraybackslash}p{0.19\linewidth}>{\raggedright\arraybackslash}Xr@{}}
\toprule
Workflow & Two evaluated claim types & P/B/F/U \\
\midrule
Phone lifecycle & Visual transition; target-state visibility & 10/10/0/0 \\
NFC launch & Physical-touch execution; phone launch visibility & 8/10/2/0 \\
Member & Phone member/authorization visibility; receiver progress/terminal state & 8/12/0/0 \\
Payment & Phone success visibility; episode-consistent cross-device outcome & 1/12/0/7 \\
\midrule
Total & 40 episodes; 80 claims & 27/44/2/7 \\
\bottomrule
\end{tabularx}
\end{table}

The collection references 126 images. Privacy review admits 118 redacted
derivatives. Eight N-device payment images are excluded because masking an
identifier would overlap the amount region; the same missing images are removed
from both arms. The Graph and Flat arms otherwise receive the same admitted
receipts, images, claim criteria, and six-label output space. Flat receives a
serialized evidence view without typed nodes, relation edges, or the
deterministic closure procedure. However, its request explicitly prohibits
using or inferring the verifier policy and includes only the policy hash. This
deviates from the pre-specified requirement of identical written-policy input.
A post-Gold input audit identified the mismatch; frozen
predictions, Gold, and scores remain unchanged. The comparison is matched at
the source-evidence and claim-criteria boundary, not at the decision-policy
boundary. Forty Flat calls and 40 Graph visual-extraction calls complete before
Gold access; 40 deterministic Graph evaluations then consume the frozen
extractions, with no retry, replacement, or tool event and with complete claim
coverage.

The post-Gold arm is a policy-transferred textual ablation, not a conventional
full-evidence LLM baseline: it receives the complete written policy and all four
deterministic verifier sources, but no typed nodes or relation edges. It compares
pipelines after transferring decision knowledge; extraction and execution
still differ, so representation alone is not isolated.

\paragraph{Frozen reference and annotation provenance.}
Two separately identified annotation returns label all 80 claims from packets
hiding predictions, internal identifiers, and the other return. Both returns
declare packet-only review without access to predictions or the other labels.
They agree on 79 claims (98.75\%; Cohen's $\kappa=0.978$); a separately
identified adjudication return resolves the disagreement. The original exported
files match the hashes recorded in frozen Gold, which also records the prediction
freeze hash before scoring.

\paragraph{Pre-specified outcomes and descriptive use.}
The pre-specified co-primary outcomes were Graph-minus-Flat PASS recall and
fixed-six-label macro-F1, each with a $-5$-point non-inferiority margin.
The missing Flat policy input prevents the planned matched-policy inference;
the frozen calculations remain available for provenance. Repeated claims,
device groups and two shared controller families are not independent sampling
units, so the field results are descriptive. We retain every episode and report
false PASS, PASS precision, blocked recall, per-class and per-workflow results,
confusion matrices, and group/family sensitivity in the artifact. The
controlled study supplies seed-level inference for the association
mechanism.

\IfFileExists{generated/natural-v8-integrated-results-v8.tex}{%
\begin{table}[htb]
\caption{Field-study claim-level results. False PASS is the error count over
53 reference non-PASS claims. Graph and Min-Flat were frozen before Gold;
Policy-Transferred Flat is a post-Gold pipeline sensitivity.}
\label{tab:natural-v8}
\centering\footnotesize
\setlength{\tabcolsep}{3pt}
\begin{tabular}{@{}lrrrrrr@{}}
\toprule
Method & Accuracy & False PASS & PASS recall & PASS precision & Macro-F1 & Blocked recall \\
\midrule
Graph & 95.00\% & 3/53 & 100.00\% & 90.00\% & 63.88\% & 90.91\% \\
Min-Flat & 65.00\% & 2/53 & 96.30\% & 92.86\% & 33.59\% & 52.27\% \\
Policy-Transferred Flat$^\dagger$ & 96.25\% & 2/53 & 100.00\% & 93.10\% & 59.02\% & 95.45\% \\
\bottomrule
\end{tabular}
\parbox{\linewidth}{\footnotesize $^\dagger$Post-Gold sensitivity; not confirmatory. Macro-F1 uses all six labels; two have zero Gold support.}
\end{table}

\paragraph{Prospective cohort observations.}
Table~\ref{tab:natural-v8} includes all 40 episodes and 80 claims.
Graph accepts all reference PASS claims while making three false PASS
judgments. The Min-Flat comparison has the policy-input mismatch described
above. Fixed-six-label macro-F1 has a ceiling of 66.67\% in this cohort because
two labels have no Gold support.

Graph's PASS precision is 27/30 (90.00\%): perfect recall does not imply
that every accepted claim is correct. Its false-PASS rate is 3/53 (5.66\%),
using reference non-PASS rather than predicted PASS as the denominator.
Table~\ref{tab:graph-confusion} accounts for all 80 claims and all four errors.
Two errors are the member episode E011's phone and receiver claims, both
reference DEVICE\_BLOCKED but predicted PASS. In NFC episode E022,
the physical-touch claim is reference DEVICE\_BLOCKED but predicted PASS;
the phone-launch visibility claim is reference DEVICE\_BLOCKED but predicted
ORACLE\_UNCERTAIN. The reference-rule sensitivity below concerns E022's
touch claim; the formal matrix remains unchanged.

\begin{table}[tb]
\caption{Graph confusion matrix against frozen Gold. Rows are reference labels;
columns are predictions. B and U denote DEVICE\_BLOCKED and ORACLE\_UNCERTAIN.
ENV\_BLOCKED and AGENT\_FAILURE have zero reference and prediction counts;
their omitted rows and columns are all zero.}
\label{tab:graph-confusion}
\centering\small
\begin{tabular}{@{}lrrrrr@{}}
\toprule
Reference & PASS & FAIL & B & U & Total \\
\midrule
PASS & 27 & 0 & 0 & 0 & 27 \\
FAIL & 0 & 2 & 0 & 0 & 2 \\
B & 3 & 0 & 40 & 1 & 44 \\
U & 0 & 0 & 0 & 7 & 7 \\
\midrule
Total & 30 & 2 & 40 & 8 & 80 \\
\bottomrule
\end{tabular}
\end{table}

\paragraph{Policy-transfer pipeline sensitivity.}
Supplying Flat with the complete written policy and deterministic verifier
sources removes the large Min-Flat gap. Both pipelines recall 100.00\% of PASS;
Policy-Transferred Flat makes two false PASS judgments versus Graph's three.
Graph has higher fixed-six-label macro-F1 (63.88\% versus 59.02\%), while
Policy-Transferred Flat has higher accuracy and blocked recall.
This post-Gold comparison characterizes pipeline sensitivity to decision
knowledge; it does not repair the planned matched-policy comparison or
establish natural Graph superiority.

\paragraph{Protocol-consistency sensitivity.}
An audit of all ten action-only touch claims found one frozen reference label
inconsistent with the prior instruction that a later capture failure does not
overturn a receipt-proven touch. With formal Gold unchanged, a separate
rule-consistency sensitivity yields 96.25\% Graph accuracy and 95.00\%
Policy-Transferred Flat accuracy, with two false PASS each.
The one-claim ordering is therefore sensitive to reference interpretation.
}{%
  \PackageError{cross-device-paper}{Missing integrated Natural V8 results}%
    {Run the Natural V8 policy-sensitivity renderer before compiling.}%
}

The diagnostic episode in Figure~\ref{fig:running-evidence} is excluded from
both evaluated cohorts.

Earlier development screens and transport-ineligible runs are retained in the
artifact history but are not pooled with either study. We report no
latency or cost advantage: the studies were designed for verdict validity, and
their scheduling and cache exposure do not support causal efficiency claims.

The supporting execution-boundary study is reported separately in
Appendix~\ref{app:supporting}. Its eight JVM cells test local timeout quarantine,
not the association benefit in R1 or natural-workflow judgment in R2.

\section{Discussion and Threats to Validity}\label{sec:discussion}

\paragraph{When explicit association is useful.}
A unique, trusted observation may need only a field selector and predicate
check. Multiple candidates with shared consumption constraints instead require
joint association: a locally plausible success path may violate another
event's binding (Figure~\ref{fig:association-conflict}). Serialized Flat shows
that computed associations can serve as an interface to a model judge; the
SQL control shows that computation need not use graphs. Applying this design
to physical records requires a matching policy grounded in the workflow.

\paragraph{Workflow composition explains the field score.}
Graph accuracy is 100\% for lifecycle and payment, and 90\% for member and
NFC launch; false-PASS counts are zero, zero, two and one, respectively.
Table~\ref{tab:workflow-claims} explains how to read these scores: only one
payment claim has reference PASS, while the other 19 are blocked or uncertain.
Accurate classification of those claims is different from completing payment.

The two labels without reference support contribute zero to the frozen
six-class macro-F1, whose ceiling in this cohort is therefore 66.67\%.
Attribution for those absent classes is not evaluated.

\paragraph{Execution conditions and claim scope.}
Two false PASS against frozen Gold share member episode E011: the member
initialization receipt reports failure, yet Graph accepts later phone and
receiver surfaces while reviewers select DEVICE\_BLOCKED. Their taxonomy lists
uninitialized devices as blocked, but does not fully specify precedence over
these local visible-state claims. This leaves two interpretations to distinguish:
a workflow prerequisite may block a dependent claim, while a standalone
page-visibility claim may still hold. The current evidence does not isolate
one interpretation as the unique cause of disagreement.

The reference-label sensitivity in Section~\ref{sec:natural-study} illustrates
the same need to distinguish an action obligation from its downstream
observation: a failed capture need not invalidate an already established touch.

\paragraph{Safety is paired with useful acceptance.}
False PASS is paired with PASS recall, macro-F1 and transport coverage.
Seed-level inference avoids treating nested claims as independent. The
association-supplied and SQL controls locate the benefit in explicit matching;
agreement with the construction reference measures contract conformance.

\paragraph{Intent authority and evidence scope.}
Generated case expectations do not independently validate business intent.
Likewise, observation scope constrains what a receipt establishes. A post-hoc
inspection of all 40 retained terminal responses finds four host queries with
matching request/response order identifiers and successful API codes: one
reports TRADE\_SUCCESS and three WAIT\_BUYER\_PAY. Query binding does not bind
an unkeyed phone or terminal image to that object. A success screen supports
its displayed state, not unobserved settlement or transaction identity.
This diagnostic does not change Gold or the reported comparisons.

\paragraph{Prediction and annotation separation.}
Predictions were frozen before Gold access, and reviewer packets hid model
outputs. The separately archived model-assisted pre-annotation is not the source
of frozen Gold. The annotation record consists of two separately identified
returns declaring packet-only review and a third adjudication return. Original
exports match the hashes in frozen Gold.

\paragraph{Dependence and external validity.}
Shared hardware and repeated executions are dependent. The field study's three
false PASS occur in one of four groups sharing two controller families; one
family has no reference PASS. Leaving out a group preserves 100\% PASS recall
but changes false PASS from zero to three, supporting descriptive interpretation.
Supporting runtime checks and historical development artifacts do not add
independent natural episodes or estimate failure prevalence.
Generalization is limited to the evaluated workflows and hardware. The
controlled matching contract isolates association behavior; transfer of its
benefit to independently validated natural associations remains untested.
The field study evaluates judgment given collected evidence, rather than
engineers' ability to author complete contracts. Supporting runtime checks
cover controlled implementation boundaries rather than provider recovery.

\paragraph{Cost and model dependence.}
An older development replay found Graph extraction calls usually slower and
more verbose than Flat. Different cache exposure, fixed within-episode arm
order and concurrent scheduling preclude a universal causal speed claim.
Both judgment and visual extraction can change with model, prompt, timeout
or source definition.

\paragraph{Relation to AdaptiTest.}
AdaptiTest scopes trace-derived experience reused in later sessions
~\cite{anonymous2026adaptitest}; this paper qualifies observations supporting a
current multi-actor verdict. Both carry scope and provenance. Our additional
object is obligation-to-observation admissibility across predicate coverage,
actor/subject/object association, temporal completeness, and claim scope.
The studies use separate episodes and results from AdaptiTest.

\section{Related Work}\label{sec:related}

\subsection{Agent Testing and Verdict Evaluation}

AndroidWorld uses application and OS-state checks, or UI validation when state
inspection is impractical~\cite{rawles2024androidworld}. MobileWorld evaluates
cross-application, tool-augmented tasks~\cite{kong2025mobileworld}; SpecOps
automates test generation through validation in real GUIs~\cite{ahmed2026specops}.

Recent mobile-testing systems enrich which actions a tester explores.
LLMDroid invokes LLM guidance when conventional exploration stagnates rather
than on every step~\cite{wang2025llmdroid}; BugHunter retrieves historical bug
reports to guide bug-aware exploration~\cite{chen2025bughunter}; and
PersonaTester injects controlled personas to reproduce diverse crowd-testing
behaviors~\cite{yu2026personatester}. Their tested knowledge improves action
selection, coverage, or behavioral diversity. Our question begins after such
actions: whether heterogeneous retained observations are eligible to support a
multi-actor business verdict, not how actions are selected.

WebTestPilot addresses implicit GUI oracles by symbolizing interface elements
and generating executable assertions over data, causal, and temporal
dependencies~\cite{teoh2026webtestpilot}. Its symbols are extracted from the
states of an observed web application. Our focus is the complementary boundary
that appears when no single application or evaluator exposes the complete
state: actions and observations belong to different physical and software
actors, and the available artifacts must first be qualified and associated
before an assertion is evaluated.

An incomplete test oracle may support only restricted inputs or
conclusions~\cite{barr2015oracle}.
CRAB evaluates agents on cross-environment tasks and uses graph-based
checkpoint evaluators for fine-grained progress measurement~\cite{xu2025crab}.
Our contract asks whether the observations behind such a checkpoint are
qualified and episode-linked for its verdict.

DevicesWorld models heterogeneous devices through a joint state while giving
the agent device-specific partial observations~\cite{li2026devicesworld}.
Its task verifier evaluates final states and generated outputs, reporting
strict success and partial progress. It therefore distinguishes local progress
from jointly satisfied goals. The difference is verifier access: our records
lack some associations and observations needed to construct a joint state.

MobileJudgeBench compares judges on 931 human-annotated trajectories from six
benchmarks~\cite{wang2026mobilejudge}. Pipeline complexity does not consistently
improve accuracy; image sampling and model backend matter (Sections 4.2 and
5.1). We distinguish trajectory success from what supplied evidence supports,
requiring a same-input, same-policy comparator.

\subsection{Distributed Oracles and Partial Observation}

Hierons studies oracle decisions for systems with physically distributed
ports, where each tester observes a local trace rather than a global
execution~\cite{hierons2012oracles}; identical local traces can represent
different global traces. We examine the corresponding qualification problem
for heterogeneous artifacts: which images, tool receipts, and business records
can support a particular obligation when identity links are partial?

PatrIoT supports integration and interoperability testing with a mixture of
physical devices, device simulations, and simulated network
infrastructure~\cite{bures2021patriot}. Its architecture includes orchestration,
reproducibility collection, and contextual reporting. We take mixed-device
execution as the setting and study the downstream oracle question of what each
artifact establishes for a business obligation.

Leotta et al. compare UML-derived Appium acceptance tests and Prolog runtime
verification in a virtualized mobile-health IoT system~\cite{leotta2019comparing}.
Their complementary UI and message-fault detection motivates our focus on
artifact-to-episode qualification before combining such evidence.

Bauer and Falcone distribute LTL obligations across local monitors without
a central trace collector~\cite{bauer2012decentralised}. Basin et al. address delayed,
reordered or lost observations using data-value quantification and
three-valued reasoning about partial knowledge~\cite{basin2019outoforder}.
These foundations supply the partial-knowledge semantics used by our contract.

Tretmans formalizes input/output conformance and quiescence under explicit
observation assumptions~\cite{tretmansQuiescence}; an arbitrary API timeout
does not supply them. Sampath et al. characterize diagnosability through
observable behavior~\cite{sampath1995diagnosability}. These results shape our
treatment of incomplete observations and unresolved outcomes.

\subsection{Trace Grounding and Evidence Acquisition}

AdaptiTest, a separate manuscript from the same authors, governs reuse of
scoped, trace-linked GUI-testing experience in later sessions
~\cite{anonymous2026adaptitest}. Its unit of decision is whether a remembered
route or observation strategy remains valid after a change of session or app
state. Our unit is a business obligation in the current execution: whether an
observation from a particular device and attempt may support that obligation.
The delta is current-run observation admissibility across heterogeneous
participants. The implemented checks separate action actor from
observation subject, bind available records to an attempt and declared object,
and preserve missing cross-device support as unresolved rather than reusing a
past instruction.

Vigil distinguishes validating a final output from validating an intermediate
artifact later consumed~\cite{li2026vigil}. Its typed events, grounded
specification terms, value bindings, and temporal patterns already address
requirement grounding and identity preservation.

Vigil assumes a trusted, complete tool-call collector (Section III-C) and uses
a closed-world fact base excluding opaque-script computation (Section IV-B).
Our adapters can omit phone reception or payment requests, so absence from the
collected trace remains uncertain. We qualify support for business test
verdicts under that observation boundary, whereas Vigil evaluates
invocation-boundary policies over its collected events.
Its model-only whole-trace baseline receives the same specification-derived
policy and trace as the other compared systems, but has no deterministic
formal semantics or localized SMT witness (Appendix B). Our policy-transfer
ablation asks a different question by supplying the verifier's decision
procedure as well as the written policy.

VAGEN combines retrospective screenshots with proactive environment
interaction under partial observability~\cite{cui2026vagen}. Its progressive
verification tools and read-only scaling (Sections 3.4--3.6.1) can obtain
additional observations. Our focus is whether those observations recover the
relation needed by the claim.

A current order query cannot reconstruct an unsaved historical request, nor can
a revisited screen recover a receive-to-launch link. Adaptive-query comparisons
must equalize permissible queries and costs.

Together, these methods motivate separating evidence selection from predicate
judgment: which retained observations support this device, attempt and claim?

\section{Conclusion}

Cross-device testing needs an answer to two questions: which observations
belong to the current obligation, and what do those observations establish?
Our executable evidence contract separates those questions and retains
unresolved support instead of promoting command completion or a success-looking
screen to business PASS.

The 12-seed controlled study isolates the first question. Atomic Flat,
association-supplied Flat and deterministic Closure produce 160, five and zero
false PASS, respectively, under the same records and business policy. Closure
improves on Atomic Flat ($p=0.00098$), while the association-supplied control
shows that explicit matching, rather than graph notation, explains most of
the gain.

On 80 natural claims, the physical-evidence implementation attains 95.00\%
accuracy with three false PASS. These observations highlight the need to
specify how execution conditions interact with each claim's observation scope.
The two studies support a practical design principle: establish admissible
support explicitly before judging its predicate, and retain the support path
so that each PASS can be audited.

\section*{Data Availability}

The local study package retains protocols, attempts, freezes, source bindings,
118 redacted images, predictions, Gold, a scorer, and runtime matrices. Local
clean-room replay reproduces the reported statistics. A separate image-free
package supports numerical replay but cannot independently verify visual
extraction or image-to-receipt binding. Raw streams, internal URLs, credentials,
annotator identifiers, and the underlying device-image collection are not
included in the public manuscript source package. Figure~\ref{fig:running-evidence}
contains redacted illustrative screenshots, distinct from that collection.
At the time of this preprint's preparation, neither the full study package nor
the prepared image-free numerical replay package is publicly hosted.
Consequently, readers cannot yet independently replay the reported results
from a public artifact. Any future release requires authorization and a separate
privacy review; no public artifact availability is claimed here.

\bibliographystyle{ACM-Reference-Format}
\bibliography{references}
\clearpage
\appendix
\section{Supporting Execution-Boundary Study}\label{app:supporting}
Eight separate JVMs cross four response conditions with legacy and strict
policies using bound execution sources and recording boundary doubles. In the
late-return condition, legacy permits one sentinel and two mapper calls and
regresses a completed status to running; strict records no late continuation
effects and keeps the state unresolved. Normal controls complete under both
policies. All eight cells preserve bindings and show no owned-thread leak.
Strict blocks continuation, not the in-flight action, and withholds cleanup
under cooperative, grace, and late returns. A wait hook controls interleavings,
not duration; strict is not verified as deployed. The complete eight-cell
matrix is retained in the study artifact.
Grace-legacy reports completed despite a resultless, timed-out action;
durable business outcomes remain unverified.

\end{document}